\documentclass[11pt]{article}
\usepackage{amsmath,amssymb,color,graphics,epsfig,cite}

\usepackage{amsfonts}
\usepackage{amssymb}

\newcommand{\hoch}[1]{$\, ^{#1}$}

\makeatletter
\@addtoreset{equation}{section}
\makeatother

\newcommand{\be}{\begin{equation}}
\newcommand{\ee}{\end{equation}}
\newcommand{\bea}{\setlength\arraycolsep{2pt} \begin{eqnarray}}
\newcommand{\eea}{\end{eqnarray}}
\newcommand{\nn}{\nonumber}

\def\ft#1#2{{\textstyle{\frac{\scriptstyle #1}{\scriptstyle #2} } }}
\def\fft#1#2{{\frac{#1}{#2}}}
\def\dfft#1#2{{\displaystyle\fft{#1}{#2}}}

\def\0{{\sst{(0)}}}
\def\1{{\sst{(1)}}}
\def\2{{\sst{(2)}}}
\def\3{{\sst{(3)}}}
\def\4{{\sst{(4)}}}
\def\5{{\sst{(5)}}}
\def\6{{\sst{(6)}}}
\def\7{{\sst{(7)}}}
\def\8{{\sst{(8)}}}
\def\sst#1{{\scriptscriptstyle #1}}
\def\oneone{\rlap 1\mkern4mu{\rm l}}
\def\ep{{\epsilon}}
\def\del{{\partial}}

\def\cH{{{\cal H}}}
\def\im{{{\rm i\,}}}
\def\R{{\mathbb R}}

\def\vp{{\varphi}}

\def\cL{{{\cal L}}}
\def\ta{{{\tilde a}}}
\def\tb{{{\tilde b}}}
\def\ba{{{\bar a}}}
\def\vep{{{\varepsilon}}}
\def\td{\tilde}
\def\wtd{\widetilde}
\def\ul{\underline}

\begin{document}

\begin{flushright}
\hfill {MI-HET-875
}\\
\end{flushright}

\begin{center}
{\large {\bf On The Consistent Supersymmetric Reduction Of Heterotic Supergravity With Geometrically Arising Yang-Mills Symmetries}}

\vspace{15pt}
{\large  C.N. Pope}

\vspace{15pt}

\hoch{1}{\it George P. \& Cynthia Woods Mitchell  Institute
for Fundamental Physics and Astronomy,\\
Texas A\&M University, College Station, TX 77843, USA}

\hoch{2}{\it DAMTP, Centre for Mathematical Sciences,
 Cambridge University,\\  Wilberforce Road, Cambridge CB3 OWA, UK}

\vspace{10pt}



\end{center}

\begin{abstract}

{\normalsize

In the paper \cite{malu}, a novel compactification of heterotic supergravity
on a warped product of $\R\times T^{1,1}$ was constructed, where
$T^{1,1}$ is a five-dimensional coset space $(SU(2)\times SU(2))/U(1)$.
It was shown that this admits a four dimensional Minkowski vacuum solution
with ${\cal N}=1$ supersymmetry, and furthermore that in the bosonic
sector there exists a remarkable fully consistent truncation in which the gauge
bosons of the $SU(2)\times SU(2)$ isometries of the $T^{1,1}$ are
retained.  In this paper, we examine this reduction further, and show that
the consistency can be extended to include the fermionic sector also.  Thus
the heterotic theory admits a consistent reduction to give an ungauged
${\cal N}=1$ supergravity coupled to $SU(2)\times SU(2)$ Yang-Mills
multiplets and a scalar multiplet.  

}

\end{abstract}

\vfill
{\small 
pope@physics.tamu.edu}
\pagebreak

\tableofcontents
\addtocontents{toc}{\protect\setcounter{tocdepth}{2}}

\section{Introduction}

  In an interesting recent paper \cite{malu}, it was observed that a 
non-trivial consistent reduction of the low-energy limit of the 
heterotic string could be constructed, by taking the six-dimensional
internal space to be a warped product $\R\times T^{1,1}$, where
$T^{1,1}$ is the five-dimensional coset space $[SU(2)\times SU(2)]/U(1)$.
Novel features of this reduction are that the associated vacuum
solution is (Minkowski)$_4\times \R\times T^{1,1}$, and that included
in the set of fields participating in the consistent reduction are
the $SU(2)\times SU(2)$ Yang-Mills fields that emerge geometrically from
the $SU(2)\times SU(2)$ isometry of the $T^{1,1}$ space.  

   Regarding the
first point, it is rather unusual for a solution involving
an internal space of non-negative scalar curvature to admit a 
Minkowski vacuum state.  A few other examples are known, however, 
including the six-dimensional Salam-Sezgin supergravity, which
admits a (Minkowski)$_4\times S^2$ vacuum solution \cite{salsez}.  As in
that case, in the new example found in \cite{malu}, the 
 (Minkowski)$_4\times \R\times T^{1,1}$ vacuum is supersymmetric.

  The second notable feature of the compactification obtained in
\cite{malu} is that the reduction including the $SU(2)\times SU(2)$ 
gauge bosons is consistent; there
is no immediately-obvious group-theoretic explanation for 
this.\footnote{Examples of reductions with a simple group-theoretic
explanations include torus reductions, and a DeWitt \cite{DeWitt} 
reduction on
a group manifold $G$ in which only the gauge bosons of one of the
two factors in the $G\times G$ isometry group of the bi-invariant metric
on $G$ are retained; in all these
cases consistency is guaranteed because one is truncating to a
subset of fields that are singlets under a transitively-acting 
subgroup of the symmetry group. By contrast, for reductions on
spheres or other coset spaces, 
first envisaged by Pauli in 1953, there is no such group-theoretic
argument for consistency, and indeed only exceptionally is the
reduction in fact a consistent one.  See, for example, 
\cite{gibpop,cvgilupo} for a more extensive discussion.}
  In this
respect it is reminiscent of the consistent reductions that arise
in certain sphere compactifications in supergravity theories, where again, 
there is no simple group-theoretic explanation for the consistency.  
Examples include the $S^7$ 
\cite{dewnic1,dewnic2,godgodnic1,godgodnic2,godgodkrunic}
and $S^4$ \cite{nasvamvan} reductions of $D=11$ supergravity, and
the $S^5$ reduction of type IIB supergravity.
In these sphere-compactification examples, a geometric
 understanding of why the
reductions are consistent has in fact been developed in more recent times, 
utilising the methods of exceptional field theory and generalised
geometry (see, for example, \cite{baghohsam,varela}).  

   A simpler, although not completely unrelated, mechanism for understanding
the consistency of certain classes of coset-space reductions was 
introduced in \cite{cvgilupo}.  In its simplest form, it is based on the
observation that a reduction of a $D$-dimensional theory 
on a coset space $K=G/U(1)$ may in certain cases 
expressible as first a (manifestly consistent) DeWitt reduction of a 
$(D+1)$-dimensional theory on the
group manifold $G$, followed by a standard Kaluza-Klein circle reduction
from $(D+1)$ to $D$ dimensions.  In those cases where such 
a $(D+1)$-dimensional origin exists, this then provides a geometric
understanding of the concomitant consistency of the Pauli reduction of the
$D$-dimensional theory on the coset $G/U(1)$.  

   It is precisely this 
construction that has been employed in \cite{malu} in order to 
construct the consistent reduction of heterotic supergravity on 
$\R\times T^{1,1}$, using the fact that $T^{1,1}$ is a coset $G/U(1)$ with
$G=SU(2)\times SU(2)$.  Another important element in the construction
in \cite{malu} is that a Scherk-Schwarz type of generalised reduction
of the heterotic supergravity yields a nine-dimensional theory with a
scalar potential term that can, at least in its bosonic sector, 
alternatively be obtained by performing a
standard Kaluza-Klein circle reduction of the ten-dimensional bosonic string 
with an added scalar potential term that has the form of a conformal 
anomaly.  Concretely, in the heterotic supergravity starting point, 
only a single $U(1)$ field is retained in the gauge sector, and the
bosonic Lagrangian thus takes the form given in eqn (\ref{d10lag}) 
below.  The generalised Scherk-Schwarz reduction to nine dimensions
is then performed using the ansatz \cite{malu}
\bea
d\hat s_{10}^2 &=& e^{2mz}\, \Big(e^{-\ft{\bar\phi}{14 a_9}} \, d\bar s_9^2
+ e^{\ft{\bar\phi}{2 a_9}}\, dz^2\Big) \,,\qquad a_9=-\sqrt{\ft89}\,,\nn\\
\hat\phi &=& -\fft{\bar\phi}{a_9} - 4 m z\,,\qquad \hat B_\2= \bar B_\2\,,
\qquad \hat A_\1= \bar A_\1\,,
\eea
where all the barred nine-dimensional fields are independent of the 10th
coordinate $z$.  In the dual description, the ten-dimensional 
starting point is the bosonic string effective action with an added
conformal anomaly term \cite{malu},
\bea
\check{\cal L}_{10} = \check R\, {\check*\oneone} -
   \ft12 {\check*d}\check\phi\wedge d\check\phi
- \ft12 e^{-\check\phi}\, {\hat* \check H_\3}\wedge \check H_\3
- 8 g^2 e^{\ft12 \check\phi}\, {\check*\oneone}\,,
\quad \check H_\3=d\check B_\2\,,
\label{checkd10lag}
\eea
which is then reduced to nine-dimensions using a standard Kaluza-Klein
reduction, with
\bea
d\check s_{10}^2 &=& e^{-\ft{\bar\phi}{14 a_9}} \, d\bar s_9^2
+ e^{\ft{\bar\phi}{2 a_9}}\, \big(dz- \ft1{\sqrt2}\, \bar A_\1\big)^2\,,\nn\\
\check\phi&=& -\fft{\bar\phi}{a_9}\,,\qquad 
 \check B_2 = \bar B_\2 + \ft1{\sqrt2}\, \bar A_\1\wedge dz\,.
\eea
(Note that in this reduction, a consistent truncation has been performed
in which the KK and winding vectors are set equal, with an associated 
truncation of one linear combination of the original ten-dimensional 
dilaton and the KK dilaton.)  In each of the two reductions the resulting
nine-dimensional theory is the same, namely \cite{malu}
\bea
\bar{\cal L}_{9} &=& \bar R\, {\bar*\oneone} -
   \ft12 {\bar*d}\bar\phi\wedge d\bar\phi
- \ft12 e^{a_9\,\bar\phi}\, {\bar* \bar H_\3}\wedge \bar H_\3
-\ft12 e^{\ft12 a_9\,\bar\phi}\, {\bar *\bar F_\2}\wedge \bar F_2
- 8 g^2 e^{\ft12 a_9\,\bar\phi}\, {\bar*\oneone}\,,\nn\\
 \check H_\3 &=&d\check B_\2 +\ft12 \bar F_\2\wedge \bar A_\1\,,
\qquad \bar F_\2=d\bar A_\1\,,
\label{d9lag}
\eea
where the constants in the two reductions are related by $g^2=8m^2$.

The ten-dimensional theory (\ref{checkd10lag}) with the conformal anomaly
term admits a (manifestly consistent)
DeWitt reduction on $G=SU(2)\times SU(2)$, with a Minkowski$_4$ vacuum. 
Thus by starting from this $G=SU(2)\times SU(2)$ DeWitt reduction, and
then reducing the ansatz in ten dimensions on the $U(1)$ Hopf fibre, it
was shown in \cite{malu}, using the techniques developed in \cite{cvgilupo},
that this provides an ansatz for the consistent
reduction of the nine-dimensional theory on the $T^{1,1}= 
[SU(2)\times SU(2)]/U(1)$ coset. Since this same nine-dimensional theory
can instead be obtained from the generalised Scherk-Schwarz reduction of
the heterotic supergravity, this provides a duality that 
allows the reduction ansatz to be lifted 
back up to ten dimensions, where it now takes the form of a
$\R\times T^{1,1}$ reduction of the heterotic supergravity.  Using the
arguments given in \cite{cvgilupo}, this reduction is necessarily a 
consistent one.

    Although the above argument provides a procedure for constructing
the consistent $\R\times T^{1,1}$ reduction ansatz for the bosonic sector
of the heterotic supergravity, it does not directly extend to the 
fermionic sector.  The reason for this is that the duality that was 
employed in the above argument relates the generalised Scherk-Schwarz
reduction of the heterotic supergravity to a standard Kaluza-Klein circle
reduction of the ten dimensional bosonic string {\it with an added 
conformal anomaly term}. Of course the ten-dimensional bosonic string
itself admits a supersymmetric extension, to ${\cal N}=1$ supergravity.
However, the inclusion of the conformal anomaly term spoils this; 
the bosonic theory with this added term cannot be 
supersymmetrised.  Thus, the derivation in \cite{malu} that yielded
a consistent $\R\times T^{1,1}$ reduction ansatz for the bosonic
sector of the heterotic supergravity cannot directly be extended to
include the fermionic sector.

  Despite the lack of a direct extension of the bosonic construction in
\cite{malu} to the fermionic sector, plausible arguments were 
presented in \cite{malu} for why a complete supersymmetric consistent
reduction of the heterotic supergravity on $\R\times T^{1,1}$ should
in fact exist. Firstly, it was shown in \cite{malu} that the
(Minkowski)$_4\times \R\times T^{1,1}$ vacuum solution of the heterotic
supergravity is supersymmetric (with ${\cal N}=1$ supersymmetry in 
four dimensions).  Secondly, it was noted that although the addition of
the conformal anomaly term to the ten-dimensional ${\cal N}=1$ supergravity
breaks the supersymmetry, the modified theory is in fact
{\it pseudo-supersymmetric} \cite{luwa,lupowa2}.  That is to say, although
the conformal anomaly term spoils the supersymmetry, it only upsets the
closure of the supersymmetry algebra beyond the quadratic order in fermion
fields.  In particular, the notion of Killing spinors in a bosonic
background configuration survives, and in fact the bosonic equations
of motion are still obtainable from the integrability conditions for
the pseudo-Killing spinors \cite{luwa,lupowa1}.

It is interesting to note that a rather similar kind of consistent reduction
was found previously in \cite{gibpop}, where it was shown that the 
six-dimensional Salam-Sezgin supergravity could be reduced on $S^2$, yielding
ungauged ${\cal N}=1$ supergravity, coupled to a scalar multiplet and
an $SU(2)$ Yang-Mills multiplet.  In that case also, the consistency of
the reduction was non-trivial, in the style of a Pauli reduction, with the
$SU(2$ Yang-Mills group deriving from the isometry group of the
compactifying $S^2$.  The derivation in \cite{gibpop} 
of the consistent reduction ansatz was carried out just by ``brute force,''
but interestingly, it was also shown in the recent paper \cite{malu} that,
at least in the bosonic sector, the reduction could be rederived more easily
by employing the same kind of techniques that were used for the
current $\R\times T^{1,1}$ example. 

It is the purpose of the present paper to give an explicit construction
of the complete ansatz for the consistent reduction of the ten-dimensional
heterotic supergravity on the internal space $\R\times T^{1,1}$, including
not only the bosonic fields as found in \cite{malu}, but also the
fermionic fields.  The resulting four-dimensional theory is ungauged
${\cal N}=1$ supergravity, coupled to a scalar multiplet and 
an $SU(2)\times SU(2)$ Yang-Mills multiplet.   The non-abelian gauge
symmetry is derived from geometric symmetries of the $T^{1,1}$ factor
in the $\R\times T^{1,1}$ internal space.

\section{Reduction of the heterotic theory}

  In this section we begin by briefly reviewing the results in \cite{malu} 
for the 
ansatz for the consistent reduction of the bosonic sector of the heterotic 
supergravity.  This will serve to establish some of our notation
and conventions, and will lead into some new material concerning the
reduction of the vielbein, which will be needed in the following section
when we construct the consistent reduction for the fermionic sector of the
heterotic theory.

  The starting point is the bosonic sector of the 
ten-dimensional heterotic theory, with the 
Yang-Mills sector specialised to the case where there is just a single $U(1)$ 
field.  The Lagrangian is
\bea
\hat \cL_{10} &=& \hat R\, {\hat*\oneone} -
   \ft12 {\hat*d}\hat\phi\wedge d\hat\phi
- \ft12 e^{-\hat\phi}\, {\hat* \hat H_\3}\wedge \hat H_\3 
-\ft12 e^{-\ft12 \hat\phi}\, {\hat* \hat F_\2}\wedge \hat F_\2 \,,
\label{d10lag}
\eea
where $\hat F_\2=d\hat A_\1$ and $\hat H_\3= d\hat B_\2 + 
\ft12 \hat F_2\wedge \hat A_\1$.  

   The theory admits a vacuum solution that is a warped product of
(Minkowski)$_4$ with $\R\times T^{1,1}$, where $T^{1,1}$ is the coset space
$[SU(2)\times SU(2)]/U(1)$ with a ``squashed'' homogeneous 
metric \cite{malu}.\footnote{The term ``squashed'' is used here to 
signify that the
metric has a different scaling factor for the $U(1)$ fibre when $T^{1,1}$
is viewed as a $U(1)$ bundle over $S^2\times S^2$, as compared with the
length of the $U(1)$ fibre for the Einstein metric on $T^{1,1}$.}

   One can view the vacuum solution as providing the basis for a
consistent dimensional reduction of the heterotic theory, which will yield
a four-dimensional ${\cal N}=1$ supergravity coupled to 
an ${\cal N}=1$ $SU(2)\times SU(2)$ Yang-Mills multiplet and a scalar
multiplet.  To see this, we first consider here the reduction ans\"atze for
the bosonic fields. 

The theory (\ref{d10lag}) has two global symmetries,
one of them being a constant shift symmetry of the dilaton, accompanied by
appropriate rescalings of the $\hat B_\2$ and $\hat A_\1$ gauge potentials.
The other is a constant trombone rescaling of the metric together with
appropriate rescalings of the gauge fields, such that the Lagrangian just
scales by an overall factor, thus implying that the equations of motion
are left invariant.  Together, these symmetries take the form
\bea
\hat g_{MN}\rightarrow \lambda^2\, \hat g_{MN}\,,\qquad
\hat B_{MN}\rightarrow \lambda^2\, e^{-\ft12 k}\, \hat B_{MN}\,,\qquad
\hat A_M\rightarrow \lambda\, e^{-\ft14k}\, \hat A_M\,,\qquad
\hat\phi\rightarrow \hat\phi +k\,.
\eea
As was observed in \cite{lalupo}, one can use global symmetries such as
these in a generalised Scherk-Schwarz like reduction scheme, by allowing
the global parameters to depend linearly on the reduction coordinate $z$.
Here, we choose the particular subgroup where $\lambda\, e^{\ft14 k}=1$,
with $k$ now taken to equal $-4m z$. 
The ten-dimensional metric is then reduced by employing the ansatz \cite{malu}
\bea
d\hat s_{10}^2= e^{2mz}\, \Big[ e^{\ft34\phi}\, ds_4^2 +
   e^{-\ft14\phi}\,(ds_5^2 + dz^2)\Big]\,,\label{metred}
\eea
where 
\bea
ds_5^2 &=& \fft1{8g^2}\, \Sigma_1^2 + \fft1{4g^2}\, (\vep^a)^2 +
       \fft1{4g^2}\, (\td\vep^\ta)^2\,,\\
\Sigma_1 &=& d\chi_1 + \cos\theta\,d\vp + \cos\td\theta\, d\td\vp
  - g \mu^\alpha\, A^\alpha - g \td\mu^\alpha \,\wtd A^\alpha\,,\\
\vep^a &=& e^a - g K^a_\alpha\, A^\alpha \,,\qquad
\td\vep^\ta = \td e^\ta - g \wtd K^\ta_\alpha\, \wtd A^\alpha\,,\label{defs1}
\eea
and where for convenience in avoiding writing $\sqrt2$ factors
it is useful to use the two related constants $g$ and $m$ by
\bea
g= 2\sqrt2\, m\,.
\eea
Here, $(\theta,\vp)$ are the coordinates of one of the $S^2$ factors in the
base, and $(\td\theta,\td\vp)$ are the coordinates for the other $S^2$,
with $e^a$ and $\td e^\ta$ being the vielbeine for the metrics on the two
unit 2-spheres:
\bea
e^a= (d\theta,\sin\theta\, d\vp)\,,\qquad 
\td e^\ta=(d\td\theta,\sin\td\theta\, d\td\vp)\,.
\eea
The quantities $\mu^\alpha$ are Euclidean embedding coordinates for
the first $S^2$ viewed as the unit sphere in $\R^3$, and similarly
$\td\mu^\alpha$ are embedding coordinates for the tilded $S^2$.  We have
\bea
\mu_1&=&\sin\theta\, \sin\vp\,,\quad \mu_2=\sin\theta\,\cos\vp\,,\quad
\mu_3=\cos\theta\,,\nn\\
\td\mu_1&=&\sin\td\theta\, \sin\td\vp\,,\quad 
\td\mu_2=\sin\td\theta\,\cos\td\vp\,,\quad
\td\mu_3=\cos\td\theta\,.
\eea
(It will often be convenient, when we assign numerical values to the $\alpha$
index on $\mu^\alpha$ or $\td\mu^\alpha$, to write the indices downstairs,
in order to avoid potential confusion with the raising of functions to powers.)
The quantities $K^a_\alpha$ and $\wtd K^\ta_\alpha$ are the vielbein components
(in the unit-sphere metrics) of the $SU(2)$ Killing vectors on the 
two 2-spheres.  Note that by a slight abuse of notation, we use the same
index $\alpha$ for labelling the three Killing vectors for each of the two 
copies of $SU(2)$ (one for each 2-sphere).  There are many useful identities
obeyed by the embedding coordinates and the Killing vectors.  An especially
useful one, from which many others follow, is that
\bea
K^m_\alpha = \ep^{mn}\,\del_n \mu^\alpha\,,\qquad
\wtd K^{\td m}_\alpha = \ep^{\td m\td n}\,\del_{\td n} \td\mu^\alpha\,.
\eea

   Note that if the $SU(2)\times SU(2)$ Yang-Mills fields are turned off,
the metric $ds_5^2$ defined in eqns (\ref{defs1}) becomes the
homogeneous non-Einstein squashed metric on $T^{1,1}$ of the 
(Minkowski)$_4\times \R\times T^{1,1}$ vacuum.  The ten-dimensional 
metric in the vacuum solution is a warped product, because of the
overall $e^{2mz}$ conformal factor in eqn (\ref{metred}).

  When we need to use explicit index values for the ten directions, our
convention will be
\bea
\hbox{4-dimensional spacetime}:&& (0,1,2,3)\nn\\
\hbox{$S^2$ 2-sphere}:&& (4,5)\nn\\
\hbox{$\wtd S^2$ 2-sphere}:&& (6,7)\nn\\
\hbox{$U(1)$ Fibre}: && 8\nn\\
\hbox{$z$ direction}: && 9\,.
\eea
The $T^{1,1}$ factor in the internal space thus corresponds to the
$(4,5,6,7,8)$ directions. Where necessary we shall use under bars on 
the numerical values of 
vielbein (flat) indices, to distinguish them from coordinate indices.

   With these preliminaries, we now present the complete bosonic 
reduction ansatz.  From the metric reduction (\ref{metred}), the reduction
ansatz for the vielbein follows, by making convenient frame choices.  
We have
\bea
\hat e^{\ul\mu} &=& e^{mz + \ft38\phi}\, e^{\ul\mu}\,,\qquad
\hat e^a= \fft1{2g}\, e^{mz - \ft18\phi}\, \vep^a\,,\qquad
\hat e^\ta= \fft1{2g}\, e^{mz - \ft18\phi}\, \td\vep^\ta\,,\nn\\
\hat e^{\ul 8} &=& \fft1{2\sqrt2\, g}\, e^{mz - \ft18\phi}\,\Sigma_1\,,\qquad
\hat e^{\ul 9}=e^{mz - \ft18\phi}\, dz\,.\label{vielans}
\eea
Note that $a=(4,5)$ and $\td a=(6,7)$.

  The ans\"atze for the remaining ten-dimensional fields are
\bea
\hat H_\3&=& H_\3 + \fft1{4g}\, (K^a_\alpha\, F^\alpha \wedge \vep^a +
                           \wtd K^\ta_\alpha\, \wtd F^\alpha\wedge\td\vep^\ta)
+\fft1{16g^2}\,(\ep_{ab}\, \vep^{a}\wedge\vep^{b}+
                \ep_{\ta\tb}\,\vep^{\ta}\wedge\vep^{\tb})
              \wedge\Sigma_1  \,,\nn\\
&& + \fft{1}{8g}\, 
  (\mu^\alpha\, F^\alpha + \td\mu^\alpha\, \wtd F^\alpha)\wedge\Sigma_1\,,
                                           \label{H3red}\\
\hat F_\2 &=& \fft{1}{4g}\, (\ep_{ab}\, \vep^a\wedge\vep^b -
                    \ep_{\ta\tb}\, \td\vep^\ta\wedge\td\vep^\tb)
   + \fft1{2}\, (\mu^\alpha\,F^\alpha -\td\mu^\alpha \wtd F^\alpha)
                             \,,\label{F2red}\\
\hat\phi &=& \ft12\phi - 4m\, z\,.\label{phired}
\eea
(The antisymmetric 
epsilon tensors $\ep_{ab}$ and $\ep_{\ta\tb}$ on the two 2-spheres
are such that $\ep_{\ul4\ul5}=1$ and $\ep_{\ul6\ul7}=1$.)  The vacuum solution
mentioned above corresponds to setting $A^\alpha=0$, $\wtd A^\alpha=0$,
$H_\3=$ and $\phi=0$ in eqns (\ref{H3red})--(\ref{phired}), together with 
taking $ds_4^2$ in eqn (\ref{metred}) to be the Minkowski metric.

  With these bosonic preliminaries and conventions established, we are now
ready to move on to the construction of the consistent reduction ansatz for
the fermionic sector.

\section{Reduction of the supersymmetry transformations}

 The supersymmetry transformation rules for the fermions in the
ten-dimensional heterotic supergravity are given by
\bea
\delta \hat \lambda &=& -\ft12 e^{-\ft14\hat\phi}\, \hat F_{AB}\,
\hat\Gamma^{AB} \,\hat\ep\,,\label{gaugino}\\
\delta\hat\chi&=& -\ft12 \hat\Gamma^M\,\del_M\hat\phi\,\hat\ep -
   \ft1{24}\, e^{-\ft12\hat\phi}\, \hat H_{ABC}\,\hat\Gamma^{ABC}\, \hat\ep
\,,\label{dilatino}\\
\delta\hat\psi_A &=& \hat D_A\,\hat\ep -\ft1{96} e^{-\ft12\hat\phi}\,
 \big(\hat\Gamma_A\,\hat\Gamma^{BCD}\, \hat H_{BCD} - 
        12\hat H_{ABC}\, \hat\Gamma^{BC}\big)\hat\ep\,.\label{gravitino}
\eea
Eqn (\ref{gaugino}) is the gaugino transformation rule, eqn (\ref{dilatino})
is the dilatino transformation rule, and eqn (\ref{gravitino}) is the
gravitino transformation rule.

  It will become necessary later to decompose the ten-dimensional 
spinors into tensor products of four-dimensional spacetime spinors and
spinors in the internal six dimensions.  Because of the structure of
the internal space it will be convenient to decompose the internal
spinors themselves as tensor products of three two-dimensional spinors.
The ten-dimensional Dirac matrices $\hat\Gamma_A$ will be decomposed as
\bea
\hat\Gamma_{\ul\mu} &=& 
         \gamma^\mu\otimes\sigma_3\otimes\sigma_3\otimes\sigma_3\,,\nn\\
\hat\Gamma_a&=& \oneone\otimes\sigma_a\otimes\sigma_3\otimes\sigma_3\,,\nn\\
\hat\Gamma_\ta&=& \oneone\otimes\oneone\otimes\sigma_\ta\otimes\sigma_3\,,\nn\\
\hat\Gamma_\ba&=& \oneone\otimes\oneone\otimes\oneone\otimes\sigma_\ba\,,
\label{Gammas}
\eea
where $\sigma_i$ are the Pauli matrices.\footnote{The notation here is a little
implicit; the index values that the index $a$, on the first $S^2$, takes
are 4 and 5, but on $\sigma_a$ these will be understood to be 1 and 2.
Similarly, for $\ta$, the actual values $\ta=(6,7)$ will be understood to
be 1 and 2 on $\sigma_\ta$.  Likewise, for $\ba=(8,9)$, these will be
understood to be 1 and 2 on $\sigma_\ba$.}  We shall give the decompositions
of the various spinor fields presently.

  Before discussing the consistent reduction of the ten-dimensional
fermion fields in detail, it is useful to summarise some of the features
that emerge if we consider the pure vacuum solution, where
the four-dimensional bosonic fields $\phi$, $A^\alpha_\mu$,
$\wtd A^\alpha_\mu$ and $B_\2$ all vanish and $g_{\mu\nu}=\eta_{\mu\nu}$.  
The vacuum
turns out to be supersymmetric, provided that $\hat\ep$ obeys certain 
projection conditions and has appropriate dependence on the coordinates
of the internal space \cite{malu}.  The
gaugino transformation rule (\ref{gaugino}) imposes one projection
condition,
\bea
(\hat\Gamma^{\ul4\ul5}-\hat\Gamma^{\ul6\ul7})\, \hat\ep=0\,,
\label{proj10}
\eea
(see eqn (\ref{proj1}) below), and the dilatino transformation
rule (\ref{dilatino}) 
imposes a second projection,
\bea
(2\hat\Gamma^{\ul8\ul9}-\hat\Gamma^{\ul4\ul5}-\hat\Gamma^{\ul6\ul7})\,
\hat\ep=0\,,\label{proj20}
\eea
(see eqn (\ref{proj2}) below).  The gravitino
transformation rule (\ref{gravitino}) then gives first-order differential
conditions on $\hat\ep$, which can be summarised as 
\bea
\hat\ep &=& e^{\ft12 m z +\ft3{16}\phi}\,\breve\ep\,,\qquad \hbox{where}
\qquad \breve\ep=  
 e^{-\ft12\chi_1\,\hat\Gamma_{\ul8\ul9}}\, \check\ep\,,\label{hatcheckep}
\eea
with the ten-dimensional spinor $\check\ep$ being entirely 
independent of the six 
internal coordinates $(\theta,\vp,\td\theta,\td\vp,\chi_1,z)$.  Since
we shall be seeing this same result in the full consistent reduction of
the ten-dimensional supersymmetry transformation rules, we
shall not give a separate derivation here for the specialisation
to the vacuum configuration.  The supersymmetry of the vacuum was also already
demonstrated in \cite{malu}.

\subsection{Gaugino transformation rule}

   Starting with the gaugino transformation rule (\ref{gaugino}), we
note from the ansatz for $\hat F_\2$ in eqn (\ref{F2red}), together
with the vielbein ansatz (\ref{vielans}), that its
vielbein components are given by
\bea
\hat F_{ab} &=& 2 g \, e^{-2mz+\ft14\phi}\, \ep_{ab}\,,\qquad
\hat F_{\ta\tb} = -2 g \, e^{-2mz+\ft14\phi}\, \ep_{\ta\tb}\,,\nn\\
\hat F_{\ul\mu\ul\nu} &=& \fft1{2}\, e^{-2mz-\ft34\phi}\, 
   (\mu^\alpha\, F^\alpha_{\ul\mu\ul\nu} -
              \td\mu^\alpha\, \wtd F^\alpha_{\ul\mu\ul\nu})\,, 
\eea
and so (\ref{gaugino}) gives
\bea
\delta\hat\lambda = 2 g\, e^{-mz +\ft18\phi}\,(\hat\Gamma^{\ul4\ul5} -
\hat\Gamma^{\ul6\ul7})\,\hat\ep -
 \fft1{4}\, e^{-mz -\ft78\phi}\, (\mu^\alpha\, F^\alpha_{\ul\mu\ul\nu}
   -\td\mu^\alpha\, \wtd F^\alpha_{\ul\mu\ul\nu})\,\hat\Gamma^{\ul\mu\ul\nu}\,
  \hat\ep\,.\label{gaugino1}
\eea

    The vacuum configuration, where the four-dimensional fields are taken
to be
$g_{\mu\nu}=\eta_{\mu\nu}$, $F^\alpha_{\mu\nu}=\wtd F^\alpha_{\mu\nu}=0$
and $\phi=0$,
should be supersymmetric.  Thus we must require that $\hat\epsilon$ satisfy
the projection condition
\bea
(\hat\Gamma^{\ul4\ul5} -\hat\Gamma^{\ul6\ul7})\,\hat\ep=0\,,\label{proj1}
\eea
as already seen in eqn (\ref{proj10}). 
We introduced the rescaled ten-dimensional supersymmetry parameter 
 $\check\ep$ in eqn (\ref{hatcheckep}). It turns out we should also then
rescale $\hat\lambda$ in the same way:
\bea
\hat\lambda = e^{-\ft12 m z -\ft3{16}\phi}\, \breve\lambda =
     e^{-\ft12 m z -\ft3{16}\phi}\, 
  e^{-\ft12\chi_1\, \hat\Gamma_{\ul8\ul9}}\, \check\lambda\,.
\label{checkeplam}
\eea
Eqn (\ref{gaugino1}) then becomes 
\bea
\delta\check\lambda= -\fft1{4}\, e^{-\ft12\phi}\, 
 (\mu^\alpha\, F^\alpha_{\ul\mu\ul\nu}
   -\td\mu^\alpha\, \wtd F^\alpha_{\ul\mu\ul\nu})\,\hat\Gamma^{\ul\mu\ul\nu}\,
  \check\ep\,.\label{gaugino2}
\eea

    Defining the reduction ans\"atze
\bea
\check\ep &=& \ep\otimes\eta\otimes\td\eta
\otimes\bar\eta\,,\label{epred}\\
\check \lambda &=&
 (\mu^\alpha\, \lambda^\alpha -\td\mu^\alpha\, \td\lambda^\alpha)\otimes
\eta\otimes\td\eta\otimes\bar\eta\,,\label{lamred}
\eea
one can then read off the supersymmetry transformations for the 
four-dimensional gaugini:
\bea
\delta\lambda^\alpha= 
 -\fft1{4}\, e^{-\ft12\phi}\, F^\alpha_{\mu\nu}\,\gamma^{\mu\nu}\,\ep\,,
\qquad
\delta\td\lambda^\alpha=
 -\fft1{4}\, e^{-\ft12\phi}\, \wtd F^\alpha_{\mu\nu}\,\gamma^{\mu\nu}\,\ep\,,
 \label{4Dgaugino}
\eea
where, for convenience, these expressions are now written using spacetime
coordinate indices rather than vielbein indices.

\subsection{Dilatino transformation rule}

   Using the ansatz (\ref{phired}) for the dilaton reduction, together with
the expressions in eqns (\ref{H3comps}) for the components of the 3-form
$\hat H_\3$, 
the ten-dimensional transformation rule (\ref{dilatino}) for the
dilatino gives
\bea
\delta\hat\chi &=& e^{-mz-\ft38\phi}\, \Big\{
 -\ft14 \hat\Gamma^{\ul\mu}\del_{\ul\mu}\phi -
      \ft1{24}\, e^{-\phi}\, H_{\ul\mu\ul\nu\ul\rho}\, 
              \hat\Gamma^{\ul\mu\ul\nu\ul\rho} +
    \fft{g}{2\sqrt2}\, e^{\ft12\phi}\,\hat\Gamma^{\ul8} \,
(2\hat\Gamma^{\ul8\ul9}- \hat\Gamma^{\ul4\ul5}-\hat\Gamma^{\ul6\ul7}) 
\label{dilatino1}\\
&& -\fft1{16}\, e^{-\ft12\phi}\, K^a_\alpha\, 
                   F^\alpha_{\ul\mu\ul\nu}\,\hat\Gamma^{\ul\mu\ul\nu a} 
 -\fft1{16}\, e^{-\ft12\phi}\, \wtd K^\ta_\alpha\, 
                   F^\alpha_{\ul\mu\ul\nu}\,\hat\Gamma^{\ul\mu\ul\nu \ta}
-\fft1{16\sqrt2}\, (\mu^\alpha \,F^\alpha_{\ul\mu\ul\nu} +
          \td\mu^\alpha \,\wtd F^\alpha_{\ul\mu\ul\nu})\,
                   \hat\Gamma^{\ul\mu\ul\nu\ul8}\Big\}\,\hat\ep\,. \nn 
\eea 
Requiring that the vacuum be supersymmetric implies the projection
condition
\bea
(2\hat\Gamma^{\ul8\ul9}- \hat\Gamma^{\ul4\ul5}-\hat\Gamma^{\ul6\ul7})
\,\hat\ep=0\,.\label{proj2}
\eea
This, together with the previous projection condition (\ref{proj1}), implies
and is implied by the conditions
\bea
\hat\Gamma^{\ul4\ul5}\,\hat\ep= \hat\Gamma^{\ul6\ul7}\,\hat\ep=
  \hat\Gamma^{\ul8\ul9}\,\hat\ep\,.\label{projall}
\eea

   Introducing first a rescaled ten-dimensional dilatino ${\breve\chi}$, 
where\footnote{Unlike in the case of the ten-dimensional 
supersymmetry parameter $\hat\ep$
and the gaugino $\hat\lambda$, we do not immediately 
define a ten-dimensional spinor $\check\chi$ that is independent
of all the internal coordinates, because of subtleties involving the
mixing of the dilatino transformation with the gaugino transformation.  See
below for a  detailed discussion.}
\bea
\hat\chi = e^{-\ft12 m z -\ft3{16}\phi}\, \breve\chi\,,\label{ulchi}
\eea
and using the rescaling of $\hat\ep$ given in the first line of 
eqn (\ref{checkeplam}), the transformation rule (\ref{dilatino1}) then becomes
\bea
\delta\breve\chi &=&
 -\fft14 \hat\Gamma^{\ul\mu}\,\del_{\ul\mu}\phi\, \breve\ep -
      \fft1{24}\, e^{-\phi}\, H_{\ul\mu\ul\nu\ul\rho}\,
              \hat\Gamma^{\ul\mu\ul\nu\ul\rho} \,\breve\ep
\label{dilatino2}\\
&& -\fft1{16}\,e^{-\ft12\phi}\, \Big[K^a_\alpha\,
  F^\alpha_{\ul\mu\ul\nu}\,\hat\Gamma^{\ul\mu\ul\nu a}
 + \wtd K^\ta_\alpha\,
              \wtd F^\alpha_{\ul\mu\ul\nu}\,\hat\Gamma^{\ul\mu\ul\nu \ta}
+\fft1{\sqrt2}\, (\mu^\alpha \,F^\alpha_{\ul\mu\ul\nu}+
          \td\mu^\alpha \,\wtd F^\alpha_{\ul\mu\ul\nu})\,
                   \hat\Gamma^{\ul\mu\ul\nu\ul8}\Big]\,\breve\ep\,. \nn
\eea

   It can be seen that the terms in the dilatino transformations
(\ref{dilatino2}) that involve $F^\alpha_{\ul\mu\ul\nu}$ are all
proportional to the four-dimensional 
gaugino transformations obtained in eqns (\ref{4Dgaugino}).
This means that the transformations in eqn (\ref{dilatino2}) are really
a combination of the four-dimensional dilatino transformation together
with an admixture of the gaugino transformations. 
The four-dimensional dilatino transformation can be disentangled
by including appropriate
subtractions of the gaugino terms when writing the reduction ansatz for
$\breve\chi$.  We find that the appropriate reduction ansatz is\footnote{Note
that in view of the projection conditions (\ref{projall}), it follows
from eqns (\ref{Gammas}) and (\ref{epred}) that $\sigma_3\eta\otimes\td\eta=
\eta\otimes\sigma_3\td\eta$, and also $\sigma_3\eta\otimes\bar\eta=
\eta\otimes\sigma_3\td\eta$ and $\sigma_3\td\eta\otimes\bar\eta=
\td\eta\otimes\sigma_3\bar\eta$.  This implies that whenever two of the
three factors $\eta$, $\td\eta$ and $\bar\eta$ have $\sigma_3$ prefactors,
they can be removed; that is $\sigma_3\eta\otimes\sigma_3\td\eta\otimes
\bar\eta= \sigma_3\eta\otimes\td\eta\otimes\sigma_3\bar\eta=
\eta\otimes\td\sigma_3\eta\otimes\sigma_3\bar\eta =
\eta\otimes\td\eta\otimes\bar\eta$.}
\bea
\breve\chi &=&
\chi\otimes\eta\otimes\td\eta\otimes\sigma_3\, 
  e^{-\ft\im2\,\chi_1\, \sigma_3}\,\bar\eta -
 \fft1{16} K^a_\alpha\,\lambda^\alpha
\otimes\sigma_a\eta\otimes\td\eta\otimes e^{-\ft\im2\,\chi_1\, \sigma_3}\,
\bar\eta\\
&& -
 \fft1{16} \wtd K^\ta_\alpha\,\td\lambda^\alpha 
 \otimes\sigma_3\eta\otimes\sigma_\ta\td\eta\otimes
e^{-\ft\im2\,\chi_1\, \sigma_3}\,\bar\eta
 -\fft1{16\sqrt2}\,(\mu^\alpha\, \lambda^\alpha +
 \td\mu^\alpha\, \td\lambda^\alpha)\otimes\eta\otimes\td\eta\otimes
\sigma_1\,e^{-\ft\im2\,\chi_1\, \sigma_3}\,\bar\eta\,.\nn
\eea

   Substituting this into eqn (\ref{dilatino1}), we can read off the 
four-dimensional supersymmetry transformation rule for the dilatino $\chi$:
\bea
\delta\chi = -\fft14 \gamma^\mu\del_\mu\phi\, \ep - \fft1{24}
e^{-\phi}\, H_{\mu\nu\rho}\, \gamma^{\mu\nu\rho}\, \ep\,.
\eea

\subsection{Gravitino transformation rule}

  The reduction of the ten-dimensional gravitino transformation rule
(\ref{gravitino}) can be simplified somewhat by introducing a new
field $\hat\Psi_A$, defined by writing
\bea
\hat\psi_A = \hat\Psi_A + \ft14 \hat\Gamma_A\,\hat\chi\,,
\eea
where $\hat\chi$ is the ten-dimensional dilatino field.  From
eqns (\ref{dilatino}) and (\ref{gravitino}), it then follows that the
ten-dimensional supersymmetry transformation of $\hat\Psi_A$ is given by
\bea
\delta\hat\Psi_A= \hat D_A\, \hat\ep + \hat P_A\,\ep  
+ \hat{\cal H}_A\, \ep\,,
\label{Psitrans2}
\eea
where 
\bea
\hat{\cal H}_A\equiv \dfft18 e^{-\ft12\hat\phi}\, 
\hat H_{ABC}\, \hat\Gamma^{BC}\,,\label{cHdef}
\eea
with its components given in 
the appendix in eqns (\ref{Hmu}) - (\ref{H9}),
and also, from eqn (\ref{phired}), that
\bea
\hat P_A\equiv \fft18\, \hat\Gamma_A\,\hat\Gamma^M\,\del_M\hat\phi =
  -\fft{m}{2}\, e^{-mz+\ft18\phi}\, \hat\Gamma_A\,\hat\Gamma^{\ul9} + 
 \fft{1}{16}\,e^{-mz-\ft38\phi}\,  \hat\Gamma_A\,    
 \hat\Gamma^{\ul\nu}\,\del_{\ul\nu}\,\phi\,.\label{Pterm}
\eea

   First, we consider the $A=\ul9$ direction of the 
supersymmetry transformation (\ref{Psitrans2}).  From
eqns (\ref{D9}), (\ref{H9}) and (\ref{Pterm}), it follows that
\bea
\delta\hat\Psi_{\ul9}=0\,,\label{Psi9trans}
\eea
after using the fact that the $z$ dependence of $\hat\ep$ is given by
eqn (\ref{epred}).  It is therefore consistent to require for
the reduction ansatz of the ten-dimensional gravitino that
\bea
\hat\Psi_{\ul9}=0\,,\qquad \hbox{and hence}\quad \hat\psi_{\ul9}= 
   \fft14\,\hat\Gamma_{\ul9}\, \hat\chi\,.
\eea

  In the $A=\ul8$ direction, the supersymmetry transformation 
$\delta\hat\Psi_A$ in eqn (\ref{Psitrans2}) gives
\bea
\delta\hat\Psi_{\ul8} = 8 m\,  e^{-mz+\ft18\phi}\,
 \Big(\fft{\del}{\del\chi_1} + \fft12 \hat\Gamma^{\ul8\ul9}\Big)\,\hat\ep
+ \fft1{16\sqrt2}\, e^{-mz-\ft78\phi}\, 
   (\mu^\alpha\, F^\alpha_{\ul\mu\ul\nu} +
 \td\mu^\alpha\, \wtd F^\alpha_{\ul\mu\ul\nu})\,\hat\Gamma^{\ul\mu\ul\nu}\,
\hat\ep\,,\label{Psi8trans1}
\eea
after making use of the already-established projection conditions
(\ref{projall}).  The supersymmetry of the vacuum therefore requires that
$\hat\ep$ satisfy the differential condition
\bea
\Big(\fft{\del}{\del\chi_1} +\fft12\,\hat\Gamma^{\ul8\ul9}\Big)\,\hat\ep
=0\,,
\eea
whose solution is
\bea
\hat\ep = e^{-\ft12\chi_1\, \hat\Gamma_{\ul8\ul9}}\,\hat\ep_0\,,
\eea
where $\hat\ep_0$ is independent of $\chi_1$.  Thus in all, we have
established that $\hat\ep$ should have the form stated previously in
eqn (\ref{hatcheckep}), namely 
\bea
\hat\ep= e^{\ft12m z+\ft3{16}\phi}\, \breve\ep= 
   e^{\ft12m z+\ft3{16}\phi}\, e^{-\ft12\chi_1\, \hat\Gamma_{\ul8\ul9}}\,
 \check\ep\,,\label{hatep}
\eea
where $\check\ep$ does not depend on $z$ or $\chi_1$. As we shall see below,
$\check\ep$ should also be independent of the coordinates 
$(\theta,\vp,\td\theta,\td\vp)$ on the two 2-spheres forming the base of the
internal $T^{1,1}$ space.
 
  With $\hat\ep$ taking the form (\ref{hatep}), the supersymmetry
transformation (\ref{Psi8trans1}) of $\hat\Psi_{\ul8}$ becomes simply
\bea
\delta\hat\Psi_{\ul8} = \fft1{16\sqrt2}\, e^{-mz-\ft78\phi}\,
   (\mu^\alpha\, F^\alpha_{\ul\mu\ul\nu} +
 \td\mu^\alpha\, \wtd F^\alpha_{\ul\mu\ul\nu})\,\hat\Gamma^{\ul\mu\ul\nu}\,
\hat\ep\,,\label{Psi8trans2}
\eea
We shall show later what this implies for the reduction ansatz for 
$\hat\Psi_{\ul8}$.

  Turning now to the $A=a$ directions, it can be seen from (\ref{Psitrans2})
that, after taking into account the projections (\ref{projall}) and the
$\chi_1$ dependence of $\hat\ep$ as in eqn (\ref{hatep}), we shall have
\bea
\delta\hat\Psi_a= 2g\, e^{-mz+\ft18\phi}\, E^m_a\,\del_m\,\hat\ep +
\fft1{16}\, e^{-mz-\ft38\phi}\, \hat\Gamma^{\ul\mu a}\, \del_{\ul\mu}\,\phi\,
\hat\ep + \fft{1}{8}\, e^{-mz-\ft78\phi}\, K^a_\alpha\, 
F^\alpha_{\ul\mu\ul\nu}\, \hat\Gamma^{\ul\mu\ul\nu}\,\hat\ep\,.
\eea
The supersymmetry of the vacuum therefore requires that $\hat\ep$ be
independent of the coordinates $(\theta,\vp)$ on the 2-sphere, and we then have
\bea
\delta\hat\Psi_a=
\fft1{16}\, e^{-mz-\ft38\phi}\, \hat\Gamma^{\ul\mu a}\, \del_{\ul\mu}\,\phi\,
\hat\ep + \fft{1}{8}\, e^{-mz-\ft78\phi}\, K^a_\alpha\,
F^\alpha_{\ul\mu\ul\nu}\, \hat\Gamma^{\ul\mu\ul\nu}\,\hat\ep\,.
\label{Psiatrans}
\eea

 The discussion for the supersymmetry transformations $\delta\hat\Psi_\ta$
go analogously, with the conclusion that $\hat\ep$ must be independent of
the coordinates $(\td\theta,\td\vp)$ on the second 2-sphere factor in the
base of the internal space, and that then we shall have
\bea
\delta\hat\Psi_\ta=
\fft1{16}\, e^{-mz-\ft38\phi}\, \hat\Gamma^{\ul\mu \ta}\, \del_{\ul\mu}\,\phi\,
\hat\ep + \fft{1}{8}\, e^{-mz-\ft78\phi}\, \wtd K^\ta_\alpha\,
\wtd F^\alpha_{\ul\mu\ul\nu}\, \hat\Gamma^{\ul\mu\ul\nu}\,\hat\ep\,.
\label{Psitatrans}
\eea

  In the spacetime directions we have from eqn (\ref{Psitrans2}) that
\bea
\delta\hat\Psi_{\ul\mu} &=& e^{-mz -\ft38\phi}\,\Big[
\del_{\ul\mu} +\fft14\hat\omega_{\ul\mu\,\ul\nu\ul\rho}\, 
\hat\Gamma^{\ul\nu\ul\rho} + \fft3{16}\, \hat\Gamma_{\ul\mu}{}^{\ul\nu}\,
\del_{\ul\nu}\, \phi + \fft1{16}\, \hat\Gamma_{\ul\mu}\,\hat\Gamma^{\ul\nu}\,
\del_{\ul\nu}\,\phi\Big]\,\hat\ep \nn\\
&&+ \fft18\,e^{-mz-\ft{11}{8}\phi}\, H_{\ul\mu\ul\nu\ul\rho}\,
\hat\Gamma^{\ul\nu\ul\rho}\,\hat\ep \,.\label{Psitransmu1}
\eea
Since the supersymmetry parameter $\hat\ep$ is then redefined in terms of
$\breve\ep$ as in eqn (\ref{hatep}) (in particular, pulling out a factor of
$e^{\ft3{16}\,\phi}$), it is nicer to rewrite (\ref{Psitransmu1}) in
terms of the supersymmetry 
parameter $\breve\ep$. We also define $\breve\Psi_{\ul\mu}$, related
to $\hat\Psi_{\ul\mu}$ by
\bea
\hat\Psi_{\ul\mu}= e^{-\ft12 m z -\ft3{16}\, \phi}\, \breve\Psi_{\ul\mu}\,,
\eea
where, in particular, $\breve\Psi_{\ul\mu}$ has no dependence on $z$.
The transformation (\ref{Psitransmu1}) then becomes
\bea
\delta\breve\Psi_{\ul\mu} &=& \Big(
\del_{\ul\mu} +\fft14\hat\omega_{\ul\mu\,\ul\nu\ul\rho}\,
\hat\Gamma^{\ul\nu\ul\rho}\Big)\,\breve\ep  + 
\fft14\, \hat\Gamma_{\ul\mu}\,\hat\Gamma^{\ul\nu}\,
(\del_{\ul\nu}\phi) \,\breve\ep 
 + \fft18\,e^{-\phi}\, H_{\ul\mu\ul\nu\ul\rho}\,
\hat\Gamma^{\ul\nu\ul\rho}\,\breve\ep 
\,.\label{Psitransmu2}
\eea

In fact for all of the vielbein components of the 
redefined ten-dimensional gravitino $\hat\Psi_A$, we can usefully 
introduce fields $\breve\Psi_A$ given by
\bea
\hat\Psi_A = e^{-\ft12 m z -\ft3{16}\phi}\, \breve\Psi_A\,,
\eea
where $\breve\Psi_A$ has no dependence on $z$.
The supersymmetry transformations (\ref{Psi9trans}), 
 (\ref{Psi8trans2}), (\ref{Psiatrans})
and (\ref{Psitatrans}) then become
\bea
\delta\breve\Psi_{\ul9} &=&0\,,\\
\delta\breve\Psi_{\ul8} &=& \fft1{16\sqrt2}\, e^{-\ft12\phi}\,
   (\mu^\alpha\, F^\alpha_{\ul\mu\ul\nu} +
 \td\mu^\alpha\, \wtd F^\alpha_{\ul\mu\ul\nu})\,\hat\Gamma^{\ul\mu\ul\nu}\,
\breve\ep\,,\label{Psi8trans3}\\
\delta\breve\Psi_a &=&
\fft1{16}\, \hat\Gamma^{\ul\mu a}\, (\del_{\ul\mu}\phi)\,
\breve\ep + \fft{1}{8}\, e^{-\ft12\phi}\, K^a_\alpha\,
F^\alpha_{\ul\mu\ul\nu}\, \hat\Gamma^{\ul\mu\ul\nu}\,\breve\ep\,,
\label{Psiatrans3}\\
\delta\breve\Psi_\ta &=&
\fft1{16}\, \hat\Gamma^{\ul\mu \ta}\, (\del_{\ul\mu}\phi)\,\breve\ep 
 + \fft{1}{8}\, e^{-\ft12\phi}\, K^a_\alpha\,
F^\alpha_{\ul\mu\ul\nu}\, \hat\Gamma^{\ul\mu\ul\nu}\,\breve\ep\,.
\label{Psitatrans3}
\eea

   Finally, the explicit four-dimensional gravitino transformation
rules can be read off from the above results, by decomposing 
the Dirac matrices and spinors in the manner described earlier.  
As with the previous results for the gaugino and dilatino transformations,
one obtains full consistent four-dimensional expressions.  That is to
say, we have shown that
the ten-dimensional transformation rules yield fully consistent
four-dimensional equations, in which non-trivial dependence of the 
ten-dimensional fermions on the coordinates of the internal manifold factorises
on both sides of each equation.  

\section{Conclusions}

  It was shown in \cite{malu} that a compactification of the ten-dimensional 
heterotic supergravity on the warped product of 
the five-dimensional coset space 
$T^{1,1}= (SU(2)\times SU(2))/U(1)$ with $\R$ gives a four-dimensional Minkowski
spacetime vacuum which is ${\cal N}=1$ supersymmetric, and that furthermore
at the bosonic level there is an associated consistent reduction, in which
the four-dimensional bosonic sector includes the gauge bosons of the
$SU(2)\times SU(2)$ factor in the isometry of $T^{1,1}$.  In this construction
the consistency of the truncation is far from obvious.  In this paper,
we have shown that the consistency of the reduction extends also to the
fermionic sector, implying that one obtains a full 
four-dimensional ${\cal N}=1$ supergravity coupled to the $SU(2)\times SU(2)
\times U(1)$ supermultiplets (plus a dilaton multiplet).  This is closely analogous to what was seen
in the case of the 2-sphere reduction of the six-dimensional Salam-Sezgin
supergravity, where the consistency of the (supersymmetric)
reduction was demonstrated
both in the bosonic and the fermionic sectors \cite{gibpop}.

   The consistency of the bosonic reduction discussed in \cite{malu} is
quite non-trivial when viewed directly as a reduction of heterotic 
supergravity.  However, as was shown in \cite{malu}, at the bosonic
level the ten-dimensional
heterotic theory is related by a duality, via a Scherk-Schwarz reduction to 
nine-dimensions, to a ten-dimensional theory that is the low-energy
limit of the bosonic string with an added conformal anomaly term 
(see eqn (\ref{checkd10lag})).  In the dual picture, the same four-dimensional
theory is simply obtained via a group-manifold reduction (of the kind first
discussed in \cite{DeWitt}), in which the consistency of the truncation
that retains the $SU(2)\times SU(2)$ of gauge bosons associated with
the left action of $SU(2)\times SU(2)$ on the group manifold is manifest.
(The argument demonstrating the consistency of the coset reduction in 
such cases was given in \cite{cvgilupo}).

An interesting question is whether the consistency of the reduction 
in the fermionic sector, which we demonstrated in this paper by a direct
construction of the reduction of the heterotic supergravity transformation
rules, could also be more simply demonstrated in the dual picture.  A 
difficulty in following this route, however, is that the ten-dimensional
bosonic theory in the dual picture (the low-energy limit of the bosonic
string with an added conformal anomaly term) does not admit a 
supersymmetrisation.  The obstacle is provided by the conformal anomaly
term.  The theory including the conformal anomaly does, however, admit a
{\it pseudo-supersymmetric} extension, of the kind discussed in 
\cite{luwa,lupowa2,lupowa1}.  That is to say, the bosonic theory with
conformal anomaly term can be supersymmetrised up to, but not including
quartic fermion terms.  (This is quite sufficient for establishing the
existence of Killing spinors in suitable purely bosonic 
backgrounds \cite{lupowa2}.)   Conceivably, the
duality between the two ten dimensional bosonic theories may in fact 
extend to the fermionic sector as well, suggesting that the notion of
pseudo-supersymmetry might acquire a more solid foundation via a duality 
relation to the heterotic supergravity.  It would be interesting to
investigate this further.

\section*{Acknowledgments}

C.N.P. is supported in part by DOE grant DE-SC0010813.

\appendix

\section{Reduction formulae}

   Some useful reduction formulae are collected here in this appendix.

   The ten-dimensional vielbein ansatz defining the dimensional
reduction is given in eqns (\ref{vielans}).  From this, 
torsion-free spin connection
$\hat\omega_{AB}$, defined
by $d\hat e^A=-\hat\omega^A{}_B\wedge \hat e^B$ and $\hat\omega_{AB}=
-\hat\omega_{BA}$, is given in
terms of the four-dimensional fields by
\bea
\hat \omega_{\ul\mu\ul\nu}&=&\omega_{\ul\mu \ul\nu}
   -\ft38 e^{-mz-\ft38\phi}\,
(\del_{\ul\mu} \phi\, \hat e_{\ul\nu}-\del_{\ul\nu}\phi\,\hat e_{\ul\mu})
  + \ft14 e^{-mz-\ft78\phi}\,(K^a_\alpha\, F^\alpha_{\ul\mu\ul\nu}\,\hat e^a
   + \wtd K^a_\alpha\, \wtd F^\alpha_{\ul\mu\ul\nu}\,\hat e^\ta)\,,\nn\\
\hat\omega_{\ul\mu a} &=& \ft18 e^{-mz-\ft38\phi}\,
                          \del_{\ul\mu}\phi\, \hat e^a +
    \ft14 e^{-mz-\ft78\phi}\, K^a_\alpha\,
                F^\alpha_{\ul\mu\ul\nu}\, \hat e^{\ul\nu}\,,\nn\\
\hat\omega_{\ul\mu \ta} &=& \ft18 e^{-mz-\ft38\phi}\,
                          \del_{\ul\mu}\phi\, \hat e^\ta +
    \ft14 e^{-mz-\ft78\phi}\, \wtd K^\ta_\alpha\,
                \wtd F^\alpha_{\ul\mu\ul\nu}\, \hat e^{\ul\nu}\,,\nn\\
\hat\omega_{\ul\mu\ul8}&=& \ft18 e^{-mz-\ft38\phi}\,
                          \del_{\ul\mu}\phi\, \hat e^{\ul8} +
\fft1{4\sqrt2}\, e^{-mz-\ft78\phi}\, (\mu^\alpha\, F^\alpha_{\ul\mu\ul\nu} +
         \td\mu^\alpha\, \wtd F^\alpha_{\ul\mu\ul\nu})\,\hat e_{\ul\nu}
\,,\nn\\
\hat\omega_{\ul\mu\ul9}&=&
\ft18 e^{-mz-\ft38\phi}\, \del_{\ul\mu}\phi\,\hat e^{\ul9} +
   m\, e^{-mz+\ft18\phi}\, \hat e_{\ul\mu}\,,\label{omred}\\
\hat\omega_{ab} &=&
-2g e^{-mz+\ft18\phi}\,\cot\theta\, \ep_{ab}\, \hat e^{\ul5}
+\fft{g}{\sqrt2}\, e^{-mz+\ft18\phi}\,\ep_{ab}\, \hat e^{\ul8}\,,\nn\\
\hat\omega_{\ta\td b} &=&
-2g e^{-mz+\ft18\phi}\,\cot\td\theta\, \ep_{\ta\td b}\, \hat e^{\ul7}
+\fft{g}{\sqrt2}\, e^{-mz+\ft18\phi}\,\ep_{\ta\td b}\, \hat e^{\ul8}\,,\nn\\
\hat\omega_{a\ul8}&=&
  \fft{g}{\sqrt2}\, e^{-mz+\ft18\phi}\, \ep_{ab}\, \hat e^b\,,\qquad
\hat\omega_{a\ul9}= m\, e^{-mz+\ft18\phi}\, \hat e^a\,,\nn\\
\hat\omega_{\ta\ul8}&=&
  \fft{g}{\sqrt2}\, e^{-mz+\ft18\phi}\, \ep_{\ta\td b}\, \hat e^{\td b}\,,
\qquad
\hat\omega_{\ta\ul9}=
  m\, e^{-mz+\ft18\phi}\, \hat e^\ta\,,\nn\\
\hat\omega_{a\td b}&=&0\,,\qquad
\hat\omega_{\ul8\ul9} = m\, e^{-m z+\ft18\phi}\, \hat e^{\ul8}\,.\nn
\eea

From these, the connection components $\hat\omega_{A\, BC}$ defined by
$\hat\omega_{BC}= \hat\omega_{A\, BC}\,\hat e^A$ can be read off.  These
are used in writing the ten-dimensional covariant derivative acting on
spinors, namely $\hat D_A = \hat E^M_A\, \del_M + 
\ft14 \hat \omega_{A\, BC} \, \hat\Gamma^{BC}$, where $M$ labels
ten-dimensional coordinate components, and $\hat E^M_A$ is the inverse
of the ten-dimensional vielbein $\hat e^A=\hat e^A_M\, dx^M$.
For future reference, we record here that the inverse vielbein
$\hat E_A= \hat E_A{}^M\, \del_M$ is given by
\bea
\hat E_{\ul\mu} &=& e^{-mz-\ft38\phi}\,\Big( E_{\ul\mu} +
   g\, A^\alpha_{\ul\mu}\, K_\alpha +
        g\, \wtd A^\alpha_{\ul\mu}\,\wtd K_\alpha\Big)\,,\nn\\
\hat E_{\ul4} &=& 2g\, e^{-m z +\ft18\phi}\,\fft{\del}{\del\theta}\,,\qquad
\hat E_{\ul5} = 2g\, e^{-m z +\ft18\phi}\,\Big(\csc\theta\,
\fft{\del}{\del\vp} -\cot\theta\, \fft{\del}{\del\chi_1}\Big)\,,\nn\\
\hat E_{\ul6} &=& 2g\, e^{-m z +\ft18\phi}\,\fft{\del}{\del\td\theta}\,,\qquad
\hat E_{\ul7} = 2g\, e^{-m z +\ft18\phi}\,\Big(\csc\td\theta\,
\fft{\del}{\del{\td\vp}} -\cot\td\theta\, \fft{\del}{\del\chi_1}\Big)\,,\nn\\
\hat E_{\ul8}&=& 2\sqrt2\,g\, e^{-m z +\ft18\phi}\, \fft{\del}{\del\chi_1}\,,
\qquad
\hat E_{\ul9}= e^{-m z +\ft18\phi}\, \fft{\del}{\del z}\,,\label{inverseviel}
\eea
where $E_{\ul\mu}= E_{\ul\mu}{}^\nu\del_\nu$ is the inverse vielbein
in the four-dimensional
spacetime metric $ds_4^2$, and
$K_\alpha = K^m_\alpha\, \dfft{\del}{\del y^m}$ and
$\wtd K_\alpha = \wtd K^{\td m}_\alpha\, \dfft{\del}{\del \td y^{\td m}}$
are the two sets of $SU(2)$ Killing vectors, with $y^m=(\theta,\vp)$ and
$\td y^{\td m}=(\td\theta,\td\vp)$.

  It is also useful to record the expressions for the vielbein components
of the ten-dimensional spinor covariant derivative 
$\hat D_A=\hat E^M_A\, \del_M + \ft14 \hat\omega_{A\, BC}\, \hat\Gamma^{BC}$.
From eqns (\ref{omred})), they are given by
\bea
\hat D_{\ul\mu}&=& e^{-mz-\ft38\phi}\,\Big[\del_{\ul\mu} 
 +\ft14 \omega_{\ul\mu\,\ul\nu\ul\rho}\, \hat\Gamma^{\ul\nu\ul\rho} 
  + g\,A^\alpha_{\ul\mu}\,K^m_\alpha\, \del_m +
  \wtd A^\alpha_{\ul\mu}\,K^{\td m}_\alpha\, \del_{\td m} 
 +\fft3{16}\,\hat\Gamma_{\ul\mu}{}^{\ul\nu}\,\del_{\ul\nu}\phi\Big]\nn\\
&&- \fft18\, e^{-mz-\ft78\phi}\, \Big[
K^a_\alpha\,F^\alpha_{\ul\mu\ul\nu}\,\hat\Gamma^{\ul\nu a} +
\wtd K^\ta_\alpha\,\wtd F^\alpha_{\ul\mu\ul\nu}\,
          \hat\Gamma^{\ul\nu \ta} 
+\fft1{\sqrt2}\, (\mu^\alpha\, F^\alpha_{\ul\mu\ul\nu} +
  \td\mu^\alpha\, \wtd F^\alpha_{\ul\mu\ul\nu})\,\hat\Gamma^{\ul\nu\ul8}\Big]
  \nn\\
&& + \fft{m}{2}\, e^{-mz +\ft18\phi}\, \hat\Gamma_{\ul\mu}{}^{\ul9}\,,
\label{Dmu}\\
\hat D_a &=&\fft1{16}\, e^{-mz-\ft38\phi}\, \hat\Gamma^{\ul\mu a}\, 
\del_{\ul\mu}\, \phi  
+\fft1{16}\, e^{-mz-\ft78\phi}\,
 K^a_\alpha\, F^\alpha_{\ul\mu\ul\nu}\, \hat\Gamma^{\ul\mu\ul\nu} \nn\\
&& + e^{-mz+\ft18\phi}\, 
 \Big[\fft{m}{2}\, \hat\Gamma^{a\ul9} -m\, \ep_{ab}\, \hat\Gamma^{b\ul8}
  + 2 g\,E^m_a\,\del_m -2g\,\cot\theta\,\delta^{\ul5}_a\, 
\big(\fft{\del}{\del\chi_1} +\fft12\, 
  \hat\Gamma^{\ul4\ul5}\big)\Big]\,,\label{Da}\\
\hat D_\ta &=&\fft1{16}\, e^{-mz-\ft38\phi}\, \hat\Gamma^{\ul\mu \ta}\, 
\del_{\ul\mu}\, \phi  
+\fft1{16}\, e^{-mz-\ft78\phi}\,
 \wtd K^\ta_\alpha\, \wtd F^\alpha_{\ul\mu\ul\nu}\, \hat\Gamma^{\ul\mu\ul\nu} 
 \nn\\
&&+ e^{-mz+\ft18\phi}\,
 \Big[\fft{m}{2}\, \hat\Gamma^{\ta\ul9} -m\, \ep_{\ta\tb}\,
        \hat\Gamma^{\tb\ul8}
  + 2 g\,E^{\td m}_\ta\,\del_{\td m} -2g\,\cot\theta\,\delta^{\ul7}_\ta\,
\big(\fft{\del}{\del\chi_1} +\fft12\, \hat\Gamma^{\ul6\ul7}\big)\Big]
\,,\label{Dta}\\
\hat D_{\ul8} &=& \fft1{16}\,e^{-mz-\fft38\phi}\, \hat\Gamma^{\ul\mu\ul8}\,
  \del_{\ul\mu} \phi +m\, e^{-mz+\ft18\phi}\,\Big(8\fft{\del}{\del\chi_1} +
\hat\Gamma^{\ul4\ul5}+\hat\Gamma^{\ul6\ul7} +\fft12\,
\hat\Gamma^{\ul8\ul9}\Big)\,,\label{D8}\\
\hat D_{\ul9} &=& \fft1{16}\, e^{-mz-\ft38\phi}\, \hat\Gamma^{\ul\mu\ul9}\,
\del_{\ul\mu}\phi + e^{-mz+\ft18\phi}\, \fft{\del}{\del z}\,,\label{D9} 
\eea
where $\del_{\ul\mu}=E^\nu_{\ul\mu}\, \del_\nu$ denotes the four-dimensional 
vielbein components of the four-dimensional spacetime derivatives, and $E^m_a$
and $E^{\td m}_\ta$ are the inverse zweibeine for the two unit 2-sphere
metrics.

  From the ansatz for $\hat H_\3$ in eqn (\ref{H3red}), together with
the vielbein reduction ansatz (\ref{vielans}), it can be seen
that its vielbein components are given by
\bea
\hat H_{\ul\mu\ul\nu\ul\rho} &=&
e^{-3mz-\ft98\phi}\, H_{\ul\mu\ul\nu\ul\rho}\,,\quad
\hat H_{\ul\mu\ul\nu a}= \fft12 e^{-3mz-\ft58\phi}\, K^a_\alpha\,
                                         F^\alpha_{\ul\mu\ul\nu}\,,\quad
\hat H_{\ul\mu\ul\nu \ta}= \fft12 e^{-3mz-\ft58\phi}\, \wtd K^\ta_\alpha\,
                                          \wtd F^\alpha_{\ul\mu\ul\nu}\,,\nn\\
\hat H_{ab\ul8} &=& \sqrt2\, g\, e^{-3mz+\ft38\phi}\, \ep_{ab}\,,\qquad
\hat H_{\ta\tb\ul8} = \sqrt2\, g\, e^{-3mz+\ft38\phi}\, \ep_{\ta\tb}\,,\nn\\
\hat H_{\ul\mu\ul\nu\ul8}&=&
        \fft1{2\sqrt2}\,e^{-3mz -\ft58\phi} \,
                    (\mu^\alpha\, F^\alpha_{\ul\mu\ul\nu} +
                          \mu^\alpha\, F^\alpha_{\ul\mu\ul\nu})\,.
                           \label{H3comps}
\eea

Using eqns (\ref{H3comps}), the components of $\cH_A$, defined in 
"eqn (\ref{cHdef}), are given by
\bea
\hat {\cal H}_{\ul\mu} &=& \fft18\, e^{-mz -\ft{11}{8}\phi}\,
H_{\ul\mu\ul\nu\ul\rho}\,\hat\Gamma^{\ul\mu\ul\nu\ul\rho}\nn\\
&& +
\fft18\,e^{-mz-\ft78\phi}\,\Big[K^a_\alpha\, F^\alpha_{\ul\mu\ul\nu}\,
\hat\Gamma^{\ul\nu a} + \wtd K^\ta_\alpha\, \wtd F^\alpha_{\ul\mu\ul\nu}\,
\hat\Gamma^{\ul\nu \ta} +\fft1{\sqrt2}\, 
(\mu^\alpha\, F^\alpha_{\ul\mu\ul\nu} + 
 \td\mu^\alpha\, \wtd F^\alpha_{\ul\mu\ul\nu})\,\hat\Gamma^{\ul\nu\ul8}\Big] 
\,,\label{Hmu}\\
\hat {\cal H}_a &=& \fft1{16} e^{-mz-\ft78\phi}\, K^a_\alpha\, 
F^\alpha_{\ul\mu\ul\nu}\, \hat\Gamma^{\ul\mu\ul\nu} +
  m\,e^{-mz+\ft18\phi}\,\ep_{ab}\,\hat\Gamma^{b\ul8}\,,\label{Ha}\\
\hat {\cal H}_\ta &=& \fft1{16} e^{-mz-\ft78\phi}\, \wtd K^\ta_\alpha\,
\wtd F^\alpha_{\ul\mu\ul\nu}\, \hat\Gamma^{\ul\mu\ul\nu} +
  m\,e^{-mz+\ft18\phi}\,\ep_{\ta\tb}\,\hat\Gamma^{\tb\ul8}
   \,,\label{Hta}\\
\hat{\cal H}_{\ul8} &=& \fft1{16\sqrt2}\, e^{-mz-\ft78\phi}\,
   (\mu^\alpha\, F^\alpha_{\ul\mu\ul\nu} +
 \td\mu^\alpha\, \wtd F^\alpha_{\ul\mu\ul\nu})\,\hat\Gamma^{\ul\mu\ul\nu}\, +
 m\, e^{-mz+\ft18\phi}\, (\hat\Gamma^{\ul4\ul5} +\hat\Gamma^{\ul6\ul7})
\,,\label{H8} \\
\hat{\cal H}_{\ul9} &=& 0\,.\label{H9}
\eea

\end{document}